\documentclass{appolb}
\usepackage{epsfig}

\begin{document}
\title{Was the Highest Energy Cosmic Ray a Photon?%
\thanks{Presented by P. Homola at the Cracow Epiphany Conference on Astroparticle Physics, Cracow, Poland, January 8-11, 2004}%
}
\author{P. Homola$^a$, M. Risse$^{a,b}$, D. G\'ora$^a$, D. Heck$^b$, H. Klages$^b$, J.~P\c{e}kala$^a$, B. Wilczy\'nska$^a$, H. Wilczy\'nski$^a$
\address{$^a$ Institute of Nuclear Physics PAS, Krak\'ow,
ul. Radzikowskiego 152, 31-342 Krak\'ow, Poland}
\address{$^b$ Forschungszentrum Karlsruhe, Institut f\"ur Kernphysik, 76021 Karlsruhe,
Germany}
}
\maketitle
\begin{abstract}
The hypothetical photonic origin of the most energetic air shower detected by the
Fly's Eye experiment is discussed. The method used for the analysis is based on 
Monte Carlo simulations including the effect of precascading of ultra-high energy (UHE) 
photons in the geomagnetic field. The application of this method to data
expected from the Pierre Auger Observatory is discussed. The importance
of complementing the southern Auger location by a northern site for UHE photon
identification is pointed out.
\end{abstract}
  
\section{Introduction}
The existence of cosmic rays of ultra-high energies (UHE), i.e. around $10^{20}$eV, is experimentally proven,
but their composition and origin are still unknown. Classic acceleration ``bottom-up'' scenarios favor hadrons
as primary cosmic rays. These scenarios require that potential accelerating sites 
should exist within several tens of Mpc from Earth. Because of the interaction
with the cosmic microwave background radiation, UHE particles from more distant objects
are not expected to reach the Earth. Due to the lack of obvious candidate sources in
our astronomical vicinity, ``bottom-up'' scenarios face serious difficulties in explaining the existence
of UHE cosmic rays. Another class of scenarios, so-called ``top-down'' models, 
generally predicts a large fraction of photons in the observable UHE cosmic-ray flux. In these scenarios
exotic physics effects are assumed including decays of supermassive ``X-particles'' which could
be produced by topological defects like cosmic strings or magnetic monopoles \cite{topdown}. 
Some of the ``top-down'' scenarios predict a reduced fraction of photons in the flux, and also certain
fraction of photons reaching the Earth's atmosphere is admitted by ``bottom-up'' scenarios. 
In any case, the identification of photon primaries, measurement of the UHE photon flux, or
specifying the upper limit for it will provide strong 
constraints on models of cosmic-ray origin.

With this motivation we attempted to analyze the highest energy air shower detected by the Fly's Eye 
experiment \cite{detector} in Utah, USA (40$^\circ$ N, 113$^\circ$ W), on 15 October 1991. 
The event parameters were finally reconstructed
in 1995 \cite{event}. The primary particle arrived at zenith angle of 43.9$^{+1.8}_{-1.3}$ deg 
and azimuth of 31.7$^{+4.2}_{-6.1}$ deg (measured counter-clockwise from East), 
its energy was 3.20$^{+0.92}_{-0.94}$ $\times$ 10$^{20}$ eV and
the atmospheric depth of shower maximum was found at 815$\pm{60}$ g/cm$^2$.
The primary particle mass was also discussed in Ref. \cite{event},
but only qualitative conclusions were drawn, namely that the shower 
profile agrees to expectations for hadron-induced events.
It could have been either a proton or a heavy nucleus, but the best fits of the observed
parameters to the expected ones were obtained for mid-size nuclei. 

The hypothetical photonic origin of the shower was discussed by Halzen et al. \cite{halzen}. 
The effect of precascading of primary gamma (preshower effect) in the geomagnetic field
was taken into account in order to find the depth of photon-induced shower maximum. The comparison of this
value to the experimental data led the authors to the conclusion that the hypothesis of the event
being a $\gamma$-ray was inconsistent with the observations.

A more accurate analysis of hypothetic photonic origin of the Fly's Eye record
event is discussed in this work. We focus on the main results that are
described in more detail in Ref. \cite{markus_fe}.
With detailed Monte Carlo simulations including 
CORSIKA \cite{corsika,lpmcorsika} and
our original code PRESHOWER \cite{preshcors} described in Section \ref{codes}, 
we obtained a set of complete photon-induced shower profiles,
compared them to the observed profile and computed the probability of the event being a photon
(Section \ref{results}).
Probabilities for different hadron primaries are also given for completeness.
Our software includes a more accurate model of the geomagnetic field than 
the one used in Ref. \cite{halzen}. The results presented in this work are also free of 
a numerical error (to be commented on later) 
present in the publication by Erber \cite{erber}. This
publication is widely cited, also in Ref. \cite{halzen}, as a standard reference 
for the cross sections necessary for computation of the cascades originated 
by UHE photons before they enter the Earth's atmosphere. 
The present, updated analysis weakens the conclusion 
given in Ref. \cite{halzen} by showing that the photon primary hypothesis 
can not be excluded.

The presented study is also an important step towards an analysis of the UHE data from the forthcoming
cosmic-ray experiments with extremely large apertures like Pierre Auger \cite{augerea} or EUSO \cite{euso}. 
Our method of analysis can be easily adopted to any
geographical position and any parameters of the primary particle. In Section \ref{auger}  
we discuss the sensitivity of the Pierre Auger Observatory to UHE photon primaries. 
In particular we compare the southern and northern site of the Observatory
with respect to their local geomagnetic conditions that influence 
the characteristics of UHE photon-induced showers in a different way. 
 
\section{Simulations}
\label{codes}
To check for a photonic origin of the Fly's Eye event, the following analysis chain is applied.
First, the propagation of UHE photons before entering the Earth's atmosphere
is simulated with a Monte Carlo code. 
This includes the proper accounting for creation of preshowers -- the effect
of precascading of UHE photons in the presence of the geomagnetic field. Thereafter, 
another Monte Carlo simulation is involved to produce extensive
air showers (EAS) induced by single UHE photons or, in cases where a preshower was
created, by the resulting bunches of less energetic particles. Then the profiles of 
such simulated showers 
are compared to the data to estimate the probability of an UHE cosmic-ray primary 
being a photon. We stress that due to shower fluctuations and measurement uncertainties, 
in general it is not possible to assign unambiguously a primary
particle type to an observed event. Thus we only estimate a probability for a photon and,
for completeness, for other primary types.

\subsection{Preshower formation}
\label{subsec-preshower}

Preshower features have been investigated by various 
authors, see for instance
[4, 8, 12-20].
Below we give only a short description of the preshower formation process.

A photon of energy above $10^{19}$~eV, in the geomagnetic field, can convert into an 
electron-positron pair before entering the atmosphere. 
The conversion probability depends on the primary photon energy $E_0$ 
and on the magnetic field component transverse to the direction of photon motion ($B_\bot$).
The resultant electrons 
subsequently lose their energy by magnetic bremsstrahlung (synchrotron radiation). 
The probability of emitting a bremsstrahlung  photon depends on the electron energy 
and also on $B_\bot$. If the energy of the emitted photon is high enough, it can create
another electron-positron pair. In this way, instead of
the primary high-energy photon, a cascade of less energetic 
particles, mainly photons and a few electrons, will enter the atmosphere. We
call this cascade a ``preshower'' since it originates and develops above the
atmosphere, i.e.~before the ``ordinary'' shower development in air.

A more detailed analysis of the preshower effect (see Ref. \cite{preshcors}) shows that other 
accompanying phenomena like deflection of $e^{+/-}$ trajectories in the magnetic field, 
influence of solar wind, time delay of particles moving slower than photons or $\gamma$
conversion in the Sun's magnetosphere
are of minor importance and can be neglected in the simulations. Therefore,
the approximation in which all the preshower particles have the same trajectory and
arrival time at the top of atmosphere can be regarded as sufficient.

\subsection{Simulation tools}

A detailed description of PRESHOWER, the code dealing with the preshower effect which 
was applied in the present analysis, is given in Ref.~\cite{preshcors}.
In brief, the geomagnetic field components are calculated according to
the International Geomagnetic Reference Field (IGRF) model~\cite{nasa}.
The primary photon propagation is started at an altitude corresponding
to five Earth's radii.
The integrated conversion probability at larger distances is sufficiently small.
The photon conversion probability is
calculated in steps of 10 km until the conversion happens or the top of the 
atmosphere (112 km) is reached.
After conversion, in steps of 1km, the resultant electrons are checked
 for bremsstrahlung emission with an adequate
probability distribution of the emitted photon energy.
A cutoff of 10$^{12}$~eV is applied for the bremsstrahlung photons,
as the influence of photons at lower energies is negligible
for the air shower evolution.
The preshower simulation is finished when the top of atmosphere is reached. 

\begin{figure}[t]
\begin{center}
\includegraphics[height=7.5cm,angle=0]{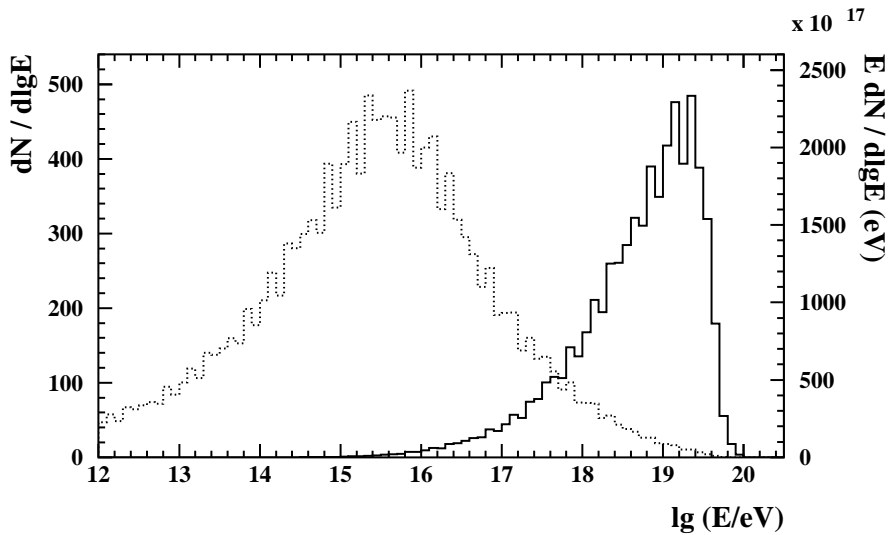}
\caption{Energy spectrum of the preshower particles (dotted line) and
spectrum weighted by energy (solid line) for the conditions of the
Fly's Eye event (average of 1000 simulation runs).}
\label{fig-espek}
\end{center}
\end{figure}

As an example, the resultant energy spectrum of the preshower particles 
for the primary photon with energy and arrival direction of the Fly's Eye record event is shown in
Figure~\ref{fig-espek} . 
We note that the photons and electrons reach the atmosphere
with energies below $10^{20}$~eV and most of the initial energy is stored
in particles of $\simeq 10^{19}$~eV. These results
don't suffer from the numerical error decreasing gamma conversion probability
which is present in the standard reference \cite{erber} on the subject.
For the Fly's Eye event parameters, calculations with this error would yield
the energy fraction contained in the particles at energies above  
$10^{19}$~eV higher by about 15\% and the number of particles in the preshower
smaller by about 20\%.

In order to analyse the properties of showers induced by UHE photons,
we combined PRESHOWER with 
a widely used air shower Monte Carlo simulation -- CORSIKA. 
Electromagnetic interactions are 
simulated in CORSIKA via the EGS4 code \cite{egs} and also 
the Landau-Pomeranchuk-Migdal (LPM) effect \cite{lpm} is taken into account
which is responsible for the increase of mean free path of electromagnetic 
particles. This increase causes a significant
delay of shower development for the electromagnetic primaries at energies 10$^{19}$ eV and larger.
Also the shower-to-shower fluctuations might be larger due to the LPM effect. 

The connection between PRESHOWER and CORSIKA is organized as follows.
PRESHOWER computes the energies and types of all the preshower particles, 
assuming for all of them the same trajectory and arrival time at the top of atmosphere.
Preshower particle data are subsequently passed to CORSIKA which processes
each particle independently, i.e. each preshower particle initiates
an atmospheric subshower and a final EAS is a superposition of these subshowers.

We want to stress that this simulation chain allows to accurately reproduce features of
UHE photon-induced showers that can be compared to the measurements. 
In particular, the shower fluctuations predicted by the MC simulations are
preserved this way. 

\section{Fly's Eye record event}
\label{results}

The arrival direction of the Fly's Eye record event makes an angle of 63$^\circ$
with the local magnetic field {\bf B}.
According to our results, this indicates that if the primary
particle were a photon, it would have produced a pair with almost 100\% probability, and the resulting
preshower would have been rather large. Now we focus on the properties
of an extensive air shower 
induced by these particles and compare them to showers induced by other primaries.

For the parameters of the Fly's Eye event, the profiles of EAS initiated
by different primary particles were produced. We obtained 1000 profiles 
of photon-induced showers with PRE\-SHO\-WER + CORSIKA. 
Shower profiles for p, C and Fe primaries were obtained with CORSIKA alone
using
two different hadronic interaction models: QGSJET~01 \cite{qgsjet01} and SIBYLL~2.1 \cite{sibyll2.1}.
1000 profiles per each primary/model were computed.
Since the Fly's Eye experiment used only the fluorescence technique of cosmic-ray detection, 
in this work we concentrated on longitudinal profiles of the showers. 

The most promising EAS feature characteristic for UHE primary photon --
the atmospheric depth of shower maximum $X_{max}$ -- was extracted from the simulated
data. For hadron primaries,
with any of the two hadronic interaction models, the $X_{max}$
values between 783 g/cm$^2$ (Fe, QGSJET 01) and 882 g/cm$^2$ (p, SIBYLL 2.1)
agree well with the measured value of $X_{max}=815\pm60$~g/cm$^2$.
For a photon primary, the value of $X_{max}=937$~g/cm$^2$ 
gives a larger, but not too large, discrepancy between the data and the expected profile 
-- about $2\sigma$.
Thus, concerning the conclusions based on the depth of shower maximum, 
neither any hadron/model combination tested, nor the photon primary hypothesis, 
can be excluded in the reconstruction of the Fly's Eye event primary type.

\begin{figure}[t]
\begin{center}
\includegraphics[height=7.5cm,angle=0]{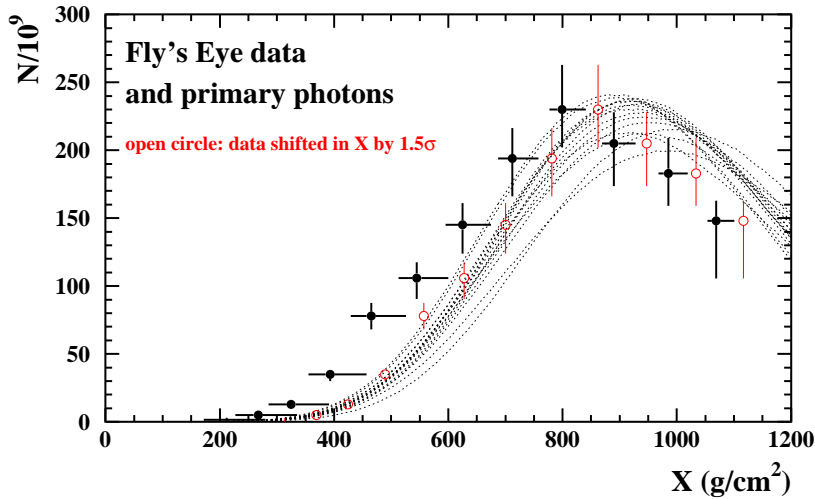}
\caption{Random subset of simulated longitudinal profiles of photon-initiated showers
compared to the Fly's Eye data as measured by the experiment (circles with vertical
and horizotal errors) and the ones shifted by $1.5\sigma$ ($X_j \rightarrow X_j +1.5 \sigma_{X_j}$)
in the direction of 
simulated profiles (open circles with only vertical errors).} 
\label{fig-profileph}
\end{center}
\end{figure}

For the statistical analysis of complete longitudinal profiles the correlation
of the atmospheric depths $X_j$ of the reconstructed data points is taken into account.
This is necessary, as the values $X_j$ emerge from a common geometry
fit to the observed signal. 
The calculation of the probability $P_i$ of each individual simulated profile 
being consistent with the measurements is based
on the reconstructed profile data as shown in Figure~\ref{fig-profileph}. 
Gaussian probability density functions for the uncertainties
$\sigma_{N_j}$ and $\sigma_{X_j}$ of the data points were assumed. Other details concerning the 
method of computing $P_i$ are discussed in Ref. \cite{markus_fe}.
The overall probability $P$
for a primary photon or other particle being consistent with the data
is obtained by averaging the probabilities $P_i$ for
the individual profiles. In this way also shower fluctuations are taken into account.
The overall probability is $P \simeq 13\%$.
This corresponds to a discrepancy between photons and data of about
1.5$\sigma$. To illustrate this, we shifted the Fly's Eye event data
by 1.5$\sigma$ and put them onto the simulated photon profiles (Fig.~\ref{fig-profileph})
and a reasonable agreement can be noted.
The quantitative analysis is published in Ref. \cite{markus_fe}.

The results obtained for the primary photon and hadron hypotheses
are summarized in Table~\ref{tab-ptot}.
\begin{table}[t]
\begin{center}
\caption{Probability $P$ of a given primary particle hypothesis to be
consistent with the observed Fly's Eye event profile and corresponding
discrepancy $\Delta$ in units of standard deviations.}
\label{tab-ptot}
\vskip 0.5 cm
\begin{tabular}{lccc}
\hline
&        &  QGSJET~01  &  SIBYLL~2.1 \\
         & photon~~~ & p~~~~~C~~~~~Fe~ & p~~~~~C~~~~~Fe \\
$P$ [\%] & 13     & ~43~~~~54~~~~53~~ & 31~~~~52~~~~54
\\
$\Delta$ [$\sigma$] & 1.5 & 0.8~~~0.6~~~0.6 & 1.0~~~0.6~~~0.6
\\
\hline
\end{tabular}
\end{center}
\end{table}
From these results a safe conclusion about the hypothetic photonic origin of
the Fly's Eye highest energy event can be drawn: 
the primary photon hypothesis, although not favored by data, cannot be excluded.
This result does not confirm the previous analysis published in Ref. \cite{halzen}.
Concerning primary hadrons, the previous conclusions are confirmed by this 
quantitative analysis. Any hadron/model
combination tested within this study is consistent with the data.

\section{Photon characteristics and the Pierre Auger Observatory}
\label{auger}
The analysis method that was applied to the Fly's Eye event
can be easily adopted for other experiments and used for data analysis on larger scale. 
As an example, the prospects for identification of photons by the Pierre Auger experiment 
are discussed.  

In Table \ref{tab-showers} collected are the simulation results of photon-induced EAS profiles
for conditions of the southern Auger Observatory in Malarg\"ue, Argentina (35.2$^\circ$S, 69.2$^\circ$W).
For different primary energies and arrival directions, full Monte Carlo simulations of photon-induced 
EAS were performed with use of PRESHOWER+CORSIKA. 
Similarly to the Fly's Eye event simulations, the strong and weak $B_\bot$ directions are defined with 
respect to the local magnetic conditions of Malarg\"ue, in
the frame where the azimuth increases counter-clockwise from East.

\begin{table}[t]
\begin{center}
\caption{Basic properties of exemplary photon-induced
showers for magnetic conditions of Malarg\"ue (35.2$^\circ$S, 69.2$^\circ$W).}
\label{tab-showers}
\begin{tabular}
{p{1.5cm}p{2.5cm}ccc} \hline \hline
$E_0$ [eV]& arrival & fraction of & $\langle X_{max} \rangle$ & 
$\langle RMS(X_{max}) \rangle$  \\ 
& direction & converted & [g/cm$^2$] & [g/cm$^2$] \\ \hline 
10$^{19.5}$ & strong $B_\bot$ & 1/50 & 1065 & 90 \\ 
10$^{20.0}$ & weak $B_\bot$ & 1/100 & 1225 & 175 \\ 
10$^{20.0}$ & strong $B_\bot$ & 91/100 & 940 & 85 \\ 
10$^{21.0}$ & weak $B_\bot$ & 100/100 & 1040 & 40 \\ 
10$^{21.0}$ & strong $B_\bot$ & 100/100 & 965 & 20 \\ \hline
\multicolumn{5}{c}{strong $B_\bot$: $\theta=53^\circ, \phi=267^\circ$; 
weak $B_\bot$: $\theta=53^\circ, \phi=87^\circ$}\\
\hline \hline

\end{tabular}\\
\end{center}
\end{table}

Analyzing the values given in Table \ref{tab-showers}, some EAS signatures
that could be helpful in identification of primary photons as cosmic rays can be listed.
First, the $X_{max}$ value of an EAS initiated by unconverted photon is 
extraordinarily deep. In this case also the fluctuations of $X_{max}$ are larger than 
in hadronic showers.
For an example, one can look at primary energy of 10$^{20}$ eV and 
the weak $B_\bot$ arrival direction, where almost no conversion of primary photons occurs and
$X_{max}(\gamma) = 1225 \pm 175$ g/cm$^2$ which is much larger than the value typical for
proton primaries of the same energy: $X_{max}(p) = 820 \pm 60$ g/cm$^2$. This signature
could allow for identification of photons on event-by-event basis.   

Another promising feature is the directional dependence of $X_{max}$ and RMS($X_{max}$).
As an example of it, consider showers of primary energies equal 10$^{20}$ eV, 
arriving from two different directions: weak $B_\bot$ and strong $B_\bot$.
Both $X_{max}$ and its fluctuations are smaller for the strong $B_\bot$ direction, for which in most
cases gamma conversion took place, than for the weak $B_\bot$, where almost all primary photons
remain unconverted and the EAS they induce have their maxima deeper in the atmosphere. 
If photons constitute a substantial fraction of UHE cosmic-ray flux, such a directional anisotropy of
$X_{max}$ and RMS($X_{max}$)
should be seen in the experimental data, provided sufficiently high statistics is available.

The other EAS feature that is characteristic only for UHE photon primaries is the small
or negative elongation rate $dX_{max}/d\log E$. For photon-induced showers
between 10$^{20}$ eV and 10$^{21}$ eV coming from the strong $B_\bot$ direction the simulated
elongation rate 25 g/cm$^2$ is much less than $\approx$ 60 g/cm$^2$ for proton or iron showers.
For events between 10$^{20}$ eV and 10$^{21}$ eV arriving from the weak $B_\bot$ direction the
elongation rate is even {\it negative} (i.e. $X_{max}$ decreases with energy). 
This is because the preshowering effect
for photons at 10$^{21}$ eV splits the initial energy into energies less than 10$^{20}$ eV, and
at this energy level, for the weak $B_\bot$ direction, almost all the primary photons remain
unconverted and they induce air showers with deeper $X_{max}$. Studies on this feature
also require large statistics of UHE events.

To have a rough feeling of how sensitive the Pierre Auger Observatory is
to the UHE photon flux, the following evaluation has been performed.
Depending on the actual high-energy particle flux and including a
duty cycle of 10$-$15\%,
the fluorescence telescopes of the Pierre Auger Observatory
are expected to record about 30$-$50 showers with primary energies exceeding
$10^{20}$~eV within a few years of data taking \cite{augerea}. If for each of these
events probability of photon primary would be on the level of $\epsilon = 5\%$, a photon fraction
in UHE cosmic-ray flux exceeding 14\% (for 30 events) to 10\% (for 50 events)
could be excluded with 95\% confidence level.
In case of the Fly's Eye event, a value of $\epsilon = 5\%$ corresponds
to reducing e.g.~the uncertainties $\sigma_{X_j}$ by a factor 1.5, which seems
well in reach for the Auger experiment.
Such un upper limit for the photon contribution to the flux would be a serious constraint for  
models of comic-ray origin.

Besides the already active detectors in Argentina, the Pierre Auger Project includes a plan 
to build an observatory in the northern hemisphere which probably will be located
close to the original Fly's Eye site. Apart from other advantages of having observatories
on two hemispheres, like for instance full sky coverage, there is also a ``pro'' argument
regarding identification of photons. 

The local geomagnetic fields differ significantly for Auger North and South both in
orientation and in strength, thus the preshower effect for primary photons is
different for the two locations. For Auger South, the local magnetic field
vector points 35$^\circ$ upwards at an azimuth of 86$^\circ$, while for Auger North
it points downwards with 66$^\circ$ at an azimuth of 75$^\circ$ (azimuth measured
counterclockwise from East). With the Auger North location being closer to the
magnetic pole, the local field strength of 0.54 Gauss is about twice the Auger
South value (0.25 Gauss).

\begin{figure}[t]
\begin{center}
\includegraphics[height=7.5cm,angle=0]{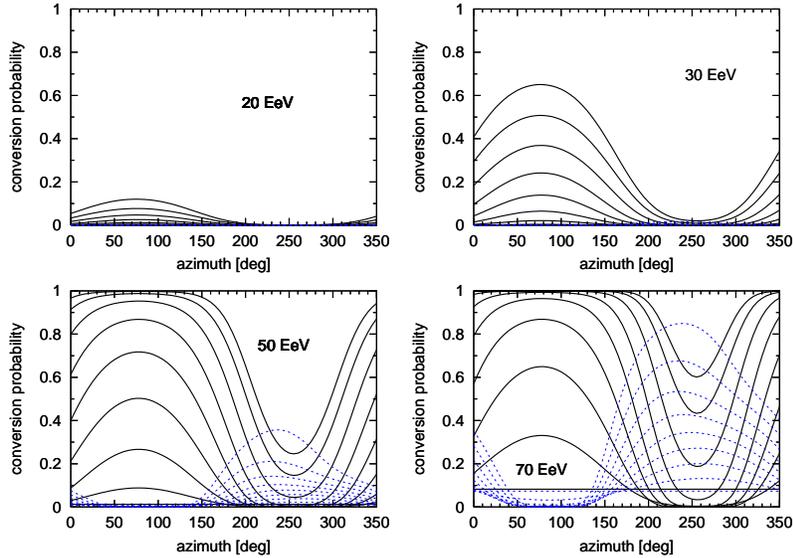}
\caption{Probability of photon conversion for different arrival directions
and four primary energies. The calculations are shown for Auger North
(solid line) and Auger South (dashed line) magnetic conditions
for the year 2005. Each curve corresponds to a different zenith angle,
from 80$^\circ$ for the uppermost curve down to 0$^\circ$ for the lowest one
in steps of 10$^\circ$. Azimuth is given counterclockwise from East for
the incoming photon.}
\label{fig-conv_maps_sn}
\end{center}
\end{figure}

As an example of the consequences on preshower formation, directional
conversion probabilities of both sites are given in Fig. \ref{fig-conv_maps_sn} for different
primary photon energies. 
In addition to the different directional dependence,
the stronger magnetic field at Auger North leads to a lower threshold energy
for the preshower onset of about 20-30 EeV, while for Auger South even at
70 EeV a large photon fraction might enter the atmosphere without conversion.

This can be used to perform a photon search in a complementary way, both in
case of presence or absence of a photon signal: at Auger South (higher
preshower threshold), a larger number of unconverted photons of higher energy
would allow better distinction from hadrons on event-by-event basis.  On the
other hand, at Auger North (lower preshower threshold), the independent photon
signature of large directional dependences in the shower observables could be
tested with larger event statistics. In particular in case of a photon signal
observed at one site, looking for a photon signature at the other site that is
expected to differ in a well-predictable way, would allow a serious cross-check
and might offer conclusive confirmation of the signal.

\section{Summary and outlook}
\label{conclusions}

Detailed investigations on the primary particle type of the record Fly's Eye shower
were performed. The focus was put on the hypothetic photonic origin of this event.
With an accurate simulation tool including preshowering and the LPM 
effects, the probability of primary particle being a photon was calculated.
The discrepancy between the simulated data and measured profile at the level of 1.5$\sigma$ 
indicates that although the primary photon hypothesis
is not favored by data, it cannot be excluded.

It is pointed out that this analysis method can be applied to any other
"fluorescence experiment", and with a generalized observable set also
to ground arrays. An application to the Pierre Auger Observatory was discussed. 
At the time of writing, the southern part of it has started data taking 
and is already the largest UHE cosmic-ray detector in the world.
Measurement of UHE photon flux or specifying the upper limit for it will give 
a serious constraint for theoretical scenarios explaining the origin of UHE cosmic rays.
From our results it appears that UHE event statistics expected during
the operation of the Auger experiment will be sufficient to give good prospects for
identification of photons as UHE cosmic-ray
primaries or to determine the upper limit of their flux.
Such an approach to identification of photons will benefit considerably from 
two observatories located in different hemispheres.

~\\

{\it Acknowledgements.}  
This work was partially supported by the Polish State Committee for
Scientific Research under grants no.~PBZ KBN 054/P03/2001 and 2P03B 11024
and by the International Bureau of the BMBF (Germany) under grant
no.~POL 99/013.
MR is supported by the Alexander von Humboldt Foundation.

\end{document}